\documentclass[a4paper,11pt]{article}
\pdfoutput=1
\usepackage{jheppub}
\usepackage{lineno}
\usepackage{upgreek}
\usepackage{float, extarrows, tikz-cd}
\usepackage{graphicx}
\usepackage{slashed}
\usepackage{tabularx,ragged2e}
\usepackage{amssymb}
\usepackage{subcaption}
\usepackage{amsmath,amssymb}
\usepackage{slashed}
\usepackage{caption}
\usepackage{xcolor}
\usepackage{dsfont}
\usepackage{verbatim}
\usepackage{mathtools, xcolor,ytableau, amsfonts,tikz}
\usepackage{graphicx}
\usepackage{physics}
\usepackage{simpler-wick}
\usepackage{microtype}
\usepackage{graphicx}
\usepackage[nobottomtitles*]{titlesec}
\newcolumntype{C}{>{\Centering\arraybackslash}X}

\numberwithin{equation}{section}

\newcommand\ov{\over}
\newcommand\p{\ensuremath{\partial}}
\newcommand{\es}[2] {\begin{equation} \label{#1} \begin{split} #2 \end{split} \end{equation}}

\def\<{\langle}
\def\>{\rangle}

\def\mop#1{\mathop{\rm #1}\nolimits}
\def\vol{\mop{vol}}

\title{Islands, Double Holography, and the Entanglement Membrane}
\author[a]{Hanzhi Jiang,}        
\author[b]{Mike Blake,}        
\author[b, c]{Anthony P. Thompson}

               \affiliation[a]{Rudolf Peierls Centre for Theoretical Physics, University of Oxford, Oxford OX1 3PU, U.K.}                                                              
               \affiliation[b]{School of Mathematics, University of Bristol, Woodland Road, Bristol BS8 1UG, U.K.} 

               \affiliation[c]{Quantum Engineering Center for Doctoral Training, University of Bristol,
                Tyndall Avenue, Bristol BS8 1FD, U.K.}

\emailAdd{hanzhi.jiang@physics.ox.ac.uk}
\emailAdd{mike.blake@bristol.ac.uk}
\emailAdd{at16718@bristol.ac.uk}
\abstract{The quantum extremal island rule allows us to compute the Page curves of Hawking radiation in semi-classical gravity. In this work, we study the connection between these calculations and the thermalisation of chaotic quantum many-body systems, using a coarse-grained description of entanglement dynamics known as the entanglement membrane. Starting from a double-holographic model of eternal two-sided asymptotically AdS$_d$ ($d>2$) black hole each coupled to a flat $d$-dimensional bath, we show that the entanglement dynamics in the late-time, large-subregion limit is described by entanglement membrane, thereby establishing a quantitative equivalence between a semi-classical gravity and a chaotic quantum many-body system calculation of the Page curve. } 

\graphicspath{{./Figures/}}
\begin{document}
\maketitle
\flushbottom

\section{Introduction}
Understanding the quantum dynamics of black holes is one of the most challenging and interesting open problems in modern theoretical physics. One particularly intriguing question is the information paradox of evaporating black holes~\cite{Hawking:1975vcx,Hawking:1976ra}. As a quantum system, one would expect the evaporation process of a black hole to be $\emph{unitary}$. Namely, the entropy of Hawking radiation should follow the Page curve~\cite{Page:1993df,Page:1993wv}, which first increases with time from zero, and then decreases when the black hole is halfway through evaporation until reaching zero. The unitarity of black hole evaporation becomes more evident with the discovery of the Anti-de Sitter/Conformal Field Theory (AdS/CFT) correspondence~\cite{Maldacena:1997re,Gubser:1998bc,Witten:1998qj}. As black holes in asymptotically AdS$_{d+1}$ spacetime are dual to CFTs in $d$-dimensions whose time evolutions are unitary, the evaporation of the black holes themselves should also be unitarity. Within the framework of AdS/CFT, important technical tools for deriving the Page curve were laid out by the Hubeny-Rangamani-Ryu-Takayanagi (RT/HRT) proposal~\cite{Ryu:2006bv,Ryu:2006ef,Hubeny:2007xt} of holographic entanglement entropy (EE) and its quantum generalisations~\cite{Faulkner:2013ana,Engelhardt:2014gca}. In the later case, one minimises the generalised entropy functional~\cite{Faulkner:2013ana} to find the codimension-2 quantum extremal surface (QES)~\cite{Engelhardt:2014gca} in the bulk. The EE is then computed by the value of the generalised entropy on the QES. Moreover, the bulk regions between the QES and the AdS boundary, known as the entanglement wedge (EW)~\cite{Czech:2012bh,Headrick:2014cta,Wall:2012uf}, are reconstructable~\cite{Almheiri:2014lwa,Dong:2016eik,Jafferis:2015del,Cotler:2017erl} from purely boundary data of the corresponding subregions. 

The derivations of Page curves of evaporating black holes in semi-classical gravity were fleshed out explicitly in~\cite{Penington:2019npb,Almheiri:2019psf}. The authors coupled the AdS black hole to a non-gravitating bath, therefore allowing it to evaporate. The key ingredient in deriving the Page curve is to include the so-called entanglement $\emph{island}$, a region in the black hole interior that is $\emph{disconnected}$ to the AdS boundary, as part of the EW of the radiation region. After the Page time, the inclusion of the island purifies the Hawking modes emitted from the black hole, thereby decreasing the entropy of radiations. 

The disconnectedness of the island to the conformal boundary of AdS appears somewhat counter-intuitive. In~\cite{Almheiri:2019hni}, inspired by the Karch-Randall-Sundrum mechanisms~\cite{Randall:1999vf,Karch:2000ct}, the author investigated the case that the bulk matters themselves have a holographic dual. Through the extra dimension, the EW of radiation becomes connected; QES gets geometrised to ordinary RT/HRT surfaces in one higher dimension. This extra dimension can be viewed as realising the idea of emergent spacetime from entanglement~\cite{Maldacena:2013xja}. 

The works~\cite{Penington:2019npb,Almheiri:2019psf,Almheiri:2019hni} focused on evaporating black holes. In~\cite{Almheiri:2019yqk}, it was pointed out that there are islands in $\emph{eternal}$ two-sided asymptotically AdS black holes as well. As the black hole is taken to be in thermal equilibrium with the heat bath, the spacetime geometry is static and the problem gets simplified. Surprisingly, the islands now lie $\emph{outside}$ the horizon. The Page curve for eternal two-sided black holes first grows, and then saturates to the coarse-grained entropy of the eternal black holes. 

For simplicity, the original works~\cite{Almheiri:2019psf,Almheiri:2019hni,Almheiri:2019yqk} considered 1+1 dimensional Jackiw-Teitelboim (JT) gravity~\cite{Jackiw:1984je,Teitelboim:1983ux,Almheiri:2014cka,Maldacena:2016upp,Engelsoy:2016xyb} coupled to CFT$_2$ matters~\cite{Calabrese:2004eu}. To test the robustness of the quantum extremal island prescription, the authors of~\cite{Almheiri:2019psy} considered higher-dimensional eternal two-sided AdS black holes, generalisating the setup in~\cite{Almheiri:2019yqk}. By employing a double holographic~\cite{Almheiri:2019hni} construction, they found that island also exists in higher dimensions outside the horizon. To be more specific, the emergence of the island after the Page time is double-holographically mapped to the $\emph{saturation}$ of the ordinary HRT surface entering the interior of the black hole~\cite{Hartman:2013qma} in one higher dimension. There are, however, more subtleties with islands in higher dimensions. For example, in a series of papers~\cite{Geng:2020qvw,Geng:2020fxl,Geng:2021hlu,Geng:2021mic}, it is pointed out that in dimension greater than three, the coupling between the black hole and the bath makes the graviton massive.\footnote{In this paper, we will calculate the entropy of radiation using double holography, where from the higher-dimensional bulk perspective we still have Einstein gravity.} 

% Islands in higher-dimensional eternal two-sided AdS black holes outside the horizon were found in~\cite{Almheiri:2019psy} using a generalisation of the setup in~\cite{Almheiri:2019yqk}. By employing a double holographic~\cite{Almheiri:2019hni} construction, \cite{Almheiri:2019psy}, the entropy of radiation, competition between ordinary HRT surfaces. 

% For example, quantum extremal islands in general dimensions were simplified and further digested in~\cite{Chen:2020uac,Chen:2020hmv,Hernandez:2020nem,Grimaldi:2022suv}. 

% The work~\cite{} presents a string theoretic derivation of the page curve using the quantum island rule. 

A question somewhat related to the dynamic of Hawking radiations is the thermalisation of a pure state in a chaotic many-body system. While the full system remains a pure state under unitary time evolutions, the reduced density matrix of a subsystem is expected to be very close to a thermal density matrix at late time~\cite{Deutsch:1991msp,Srednicki:1994mfb,Rigol:2007juv}. Such a thermalisation process of the reduced density matrix can be characterised by its entanglement entropy. In Page's original calculation~\cite{Page:1993wv}, the black hole also collapses from a pure state; the fact that EE of a subregion and its complement is equal plays a crucial role. As black holes are maximally chaotic~\cite{Shenker:2013pqa,Maldacena:2015waa}, it would be tempting to conjecture that the quantum extremal island prescription is a generic feature of entanglement dynamics in chaotic many-body systems. Indeed, it was shown~\cite{Liu:2019svk,Liu:2020gnp,deBoer:2023axh} that Page curves can be derived from various chaotic many-body systems as well. Most relevantly, the work~\cite{Blake:2023nrn} calculated the Page curve from a coarse-grained model describing entanglement dynamics in a wide range of chaotic many-body systems known as the entanglement membrane~\cite{Jonay:2018yei}. Nevertheless, all these calculations are $\emph{qualitatively}$ inspired by the semi-classical gravity results. A $\emph{quantitative}$ connection between a gravity and a chaotic many-body system computation of the Page curve remains absent. 

The aim of this paper is to work out an explicit equivalence between the two aforementioned perspectives of the Page curve. Inspired by~\cite{Blake:2023nrn}, we will focus on entanglement membrane. Originally developed from random unitary circuits~\cite{Nahum:2016muy,Jonay:2018yei,Zhou:2018myl}, membrane theory can also be derived from Floquet circuits~\cite{Zhou:2019pob}, generalised dual-unitary circuits~\cite{Rampp:2023vah}, Brownian Hamiltonians~\cite{Vardhan:2024wxb}, and importantly, holographic gauge theories~\cite{Mezei:2018jco,Mezei:2019zyt,Jiang:2024tdj}. It is the last perspective that allows us to quantitatively map the Page curve calculations in semi-classical gravity~\cite{Almheiri:2019psy,Geng:2020qvw} to those in chaotic many-body systems. 

% \footnote{As we will see, one needs to adjust the parameters in~\cite{Almheiri:2019psy,Geng:2020qvw} for the resulting theories to be applicable for entanglement membrane~\cite{Mezei:2018jco}. } 

The organisation of the paper is as follows: in section~\ref{MembraneReview}, we review entanglement membrane, focusing on the holographic point of view~\cite{Mezei:2018jco,Mezei:2019zyt,Jiang:2024tdj}. In section~\ref{DoubleHoloSetup}, we review the setup of our double holographic model derived from~\cite{Geng:2020qvw,Almheiri:2019psy}. Section~\ref{PageCurveMembrane} is the core of this paper. In this section, we first find the relevant HRT surfaces computing the Page curve in the aforementioned double holographic model,\footnote{With some adjustments to the parameters so that membrane theory can be applied. } and then use them to derive an entanglement membrane theory. Comparison between~\cite{Blake:2023nrn} from purely a chaotic many-body system point of view is made in  subsection~\ref{BTreview}. In Appendix~\ref{JoiningQuench}, we provide preliminary evidence that membrane theory for joining quench~\cite{Mezei:2019zyt} can be applied to studies of Page curves for evaporating black holes via double holography before the Page time.

\section{Membrane Theory Review}\label{MembraneReview}
\begin{figure}[htbp]
\centering
\includegraphics[width=.5\textwidth]{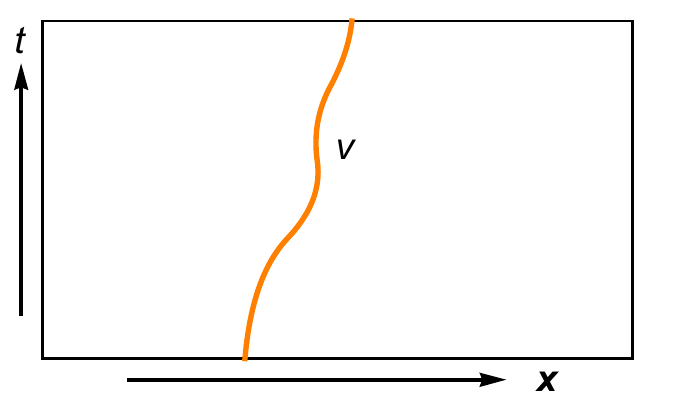}
\qquad
\caption{Cartoon of a minimal membrane (orange) extending in spacetime in $(1+1)$-dimensions. This figure represents entropy dynamics in a quench, where the top time slice is  $A(t)$, and the bottom time slice represents the initial state. \label{fig:MembraneCartoon}}
\end{figure}

In this section we review the basic elements of entanglement membrane. In membrane theory, the time-dependent entanglement entropy is computed by the action
\es{MinMemb}{
S=\min_{x(\xi)} \int d^{d-1}\xi\, \sqrt{\abs{\gamma}}\, {s_\text{th}\, {\cal E}(v)\over \sqrt{1-v^2}}\,,
}
minimised over all membrane shapes obeying some given boundary conditions, see Fig.~\ref{fig:MembraneCartoon}. In the above formula $s_\text{th}$ is the coarse-grained entropy density, $\xi$ and $x$ are the membrane worldvolume and spacetime coordinates, respectively, $\gamma$ is the induced metric of the membrane worldvolume, $v$ is the local transverse velocity of the membrane, and ${\cal E}(v)$ is the membrane tension function depending on the microscopic details of the many-body system. In a quench setup, the membrane connects the time slice of the boundary of the subregion $A(t)$ and the time slice of the initial state, see Fig.~\ref{fig:MembraneCartoon}.

Membrane theory of entanglement dynamics can be derived from holographic gauge theories by studying the HRT surfaces in dual gravitational theories~\cite{Mezei:2018jco}. Throughout this paper we will focus on asymptotically AdS$_{d+1}$  black brane in infalling coordinates:
\begin{equation}
ds^2 = \frac{1}{z^2}\left(-a(z)du^2 -{2\ov b(z)}dudz+dx^2+ d\vec{y}_{d-2}^2\right)\label{blackBraneInf}
\end{equation}
where the conformal boundary is at $z=0$, and the horizon is at $z_+=1$; therefore $a(1)=0$. The  AdS$_{d+1}$ asymptotics requires $a(0)=b(0)=1$. For planar AdS$_{d+1}$-Schwarzschild black brane $a(z)=1-z^d,\, b(z)=1$. 

% \begin{figure}[htbp]
% \centering
% \includegraphics[width=.365\textwidth]{Figures/DispHalfSpaceAdSSch.pdf}
% \qquad
% \caption{Penrose diagram of the HRT surface computing the time-dependent entanglement entropy in planar AdS$_{d+1}$-Schwarzschild geometry with entangling subregion taken to be half the space on both side of the thermofield double. The $\pm X$ displacement in the $x$ direction is perpendicular to the plane of the paper. \HJ{Add special extremal slice here}\label{fig:DispHalfSpaceAdSSch}}
% \end{figure}

\begin{figure}[htbp]
\centering
\includegraphics[width=.4\textwidth]{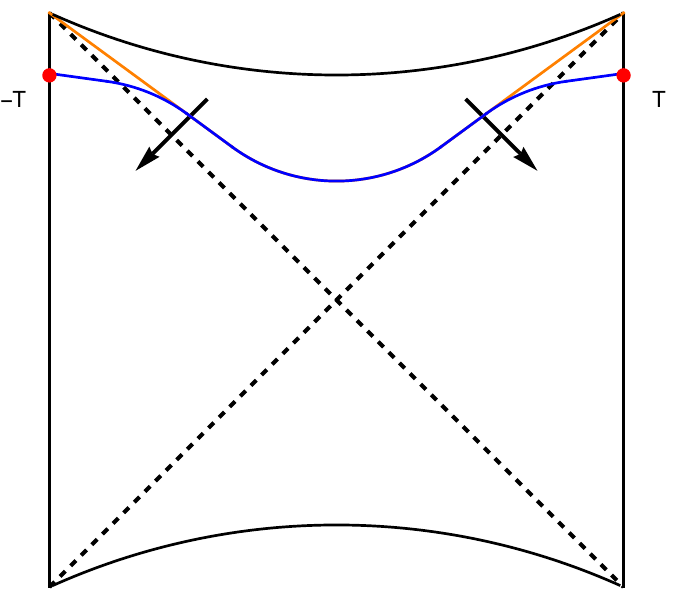}\hspace{5mm}
\includegraphics[width=.4\textwidth]{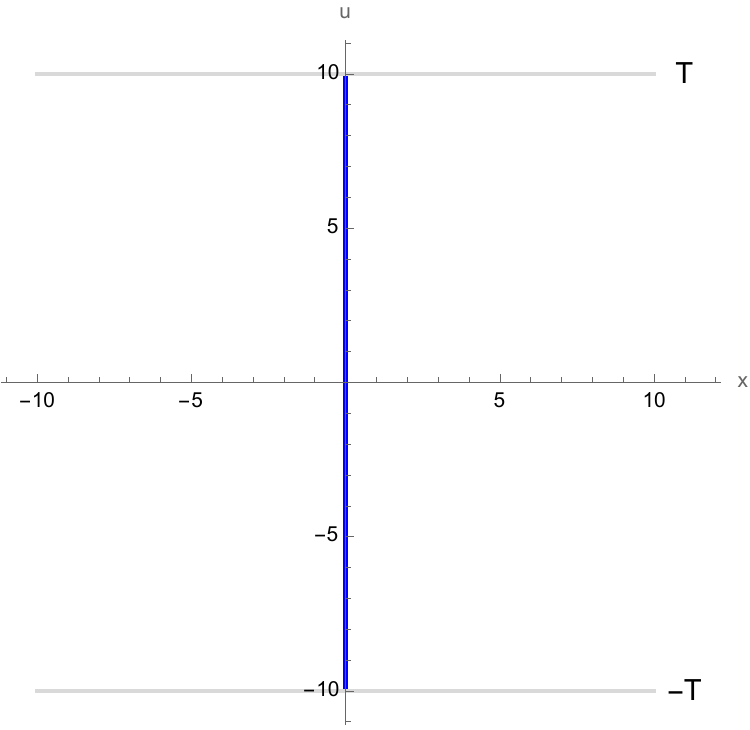}
\qquad
\caption{$\emph{Left}$: Penrose diagram of the HRT surface (blue) computing the time-dependent entanglement entropy for half-spaces in planar AdS$_{d+1}$-Schwarzschild geometry. At late time, the HRT surface gets stuck at a special extremal slice $z=z_*$ (orange) in the interior. The arrows denote constant infalling time that we project the HRT surface along. $\emph{Right}$: Minimal membrane stretching between $t=-T$ and $t=T$ with $x=0$. \label{fig:DispHalfSpaceAdSSch}}
\end{figure}

We would like to determine the two-sided entanglement entropy with strip entangling subregions extending in the $\vec{y}_{d-2}$ directions at late-time. We parametrise the HRT surface as $x(u),\, z(u)$, and obtain the area functional from~\eqref{blackBraneInf}:
\begin{equation}
\label{AreaFunct}
\begin{aligned}
S &= s_\text{th} \vol(\p A) \int du  \frac{\sqrt{Q}}{z^{d-1}}\,, \qquad Q\equiv \dot x^2-a(z)-{2\ov b(z)}\dot z
\end{aligned}
\end{equation}
where $^\cdot \equiv\frac{d}{du}$. To simplify the calculations, we implement the following self-consistent scaling Ansatz~\cite{Mezei:2018jco}
\es{CoordScaling}{
u\rightarrow \Lambda u, \hspace{0.5cm} x \rightarrow \Lambda x, \hspace{0.5cm} z\rightarrow z\,,
}
for $\Lambda\gg1$. In this scaling the $\dot z$ term can be dropped from $Q$ in \eqref{AreaFunct}. Then the $z$ equation of motion becomes algebraic~\cite{Mezei:2018jco}
\es{projectionz}{
v^2\equiv \dot x^2  = a(z) - \frac{z a'(z)}{2(d-1)} \equiv c(z)\,;
}
For example, for planar AdS$_{d+1}$-Schwarzchild black brane, we have $c(z)=1-{d-2\ov 2(d-1) }\, z^d$. The function $c(z)$ is monotonically decreasing and positive in the range $1\leq z \leq z_*$, where $z_*$ is the first zero of $c(z)$~\cite{Mezei:2018jco}. Then $z$ in this range can be solved for in terms of $v^2$; in other words, we have integrated out the radial bulk $z$ degree of freedom. The resulting reduced action is the membrane theory introduced in \eqref{MinMemb}:
\begin{equation}
\label{AreaFunct2}
\begin{aligned}
S &= s_\text{th} \vol(\p A) \int du \,  \mathcal{E}(v)\,, \qquad
\mathcal{E}(v) =\left. \sqrt{\frac{-a'(z)}{2(d-1)z^{2d-3}}}\right|_{z=c^{-1}(v^2)}\,.
\end{aligned}
\end{equation}
where the membrane tension $\mathcal{E}(v)$ is specified by the interior geometry of the black brane. The bulk infalling time $u$ plays the role of boundary theory time.\footnote{Notice that at the AdS boundary, infalling and boundary time are the same.} As we have integrated out the $z$ degree of freedom in obtaining~\eqref{AreaFunct2}, we can under the geometric meaning of $\mathcal{E}(v)$ as the $\emph{projection}$ of the bulk HRT surface to the boundary along constant infalling time, see Figure~\ref{fig:DispHalfSpaceAdSSch} for an example of the $v=0$ membrane.\footnote{For two-sided setups, we also need to $\emph{glue}$ the two projections on both sides at $u=0$. In general, membranes with strip entangling subregions are straight lines with slope $v\in[0,v_B)$. For spherical entangling subregions~\cite{Mezei:2018jco}, the membrane shapes are more complicated.} 

To further present the properties the membrane tension function $\mathcal{E}(v^2)$ obeys, we first introduce two important velocities characterising the spread of quantum information: 
\begin{itemize}
    \item The entanglement velocity $v_E$~\cite{Hartman:2013qma,Liu:2013iza,Liu:2013qca} determines the speed at which entanglement grows linearly
\begin{align}
    v_E=\sqrt{-\frac{a(z)}{z^{2(d-1)}}}\Big|_{z=z_*}\,,\label{vE}
\end{align}
where $z_*$  maximises \eqref{vE} the expression under the square root. $z_*$ is the deepest point in the black hole interior the HRT surface can reach behind the horizon in the scaling regime~\eqref{CoordScaling}.\footnote{For planar AdS-Schwarzschild black brane, $z_*=\Big(\frac{2(d-1)}{d-2}\Big)^{1/d}$~\cite{Hartman:2013qma,Liu:2013iza,Liu:2013qca}. When $d=2$, i.e. in the BTZ black brane, $z_*=\infty$. } 
    \item The butterfly velocity $v_B$~\cite{Roberts:2014isa} is related to the out-of-time order correlator (OTOC) 
\begin{align}
    v_B=\sqrt{-\frac{a'(1)}{2(d-1)}}\,.\label{vB}
\end{align}
\end{itemize}
In general, $v_E\leq v_B$ due to the constraint from null energy condition~\cite{Mezei:2016zxg}. The membrane tension function $\mathcal{E}(v)$ obeys the following constraints~\cite{Jonay:2018yei,Mezei:2018jco}: 
\begin{align}
    \mathcal{E}(0)=v_E && \mathcal{E}'(0)=0 && \mathcal{E}(v_B)=v_B && \mathcal{E}'(v_B)=1 && \mathcal{E}'(v)>0 && \mathcal{E}''(v)>0\,. \label{constraints}
\end{align}
Thus, $\mathcal{E}(v)$ is an even function, monotonically increasing and convex for $0\leq v <v_B$. 

For 2d CFT, the membrane theory becomes degenerate: we have $c(z)=1$ and $v_E=v_B=\mathcal{E}(v)=1$. In~\cite{Jiang:2024tdj}, it was shown that in the 2d CFT case, we need to generalise membrane theory by introducing an additional degree of freedom on the membrane.\footnote{Unlike ordinary membrane theory~\cite{Mezei:2018jco}, generalised membrane theory~\cite{Jiang:2024tdj} does not have a random unitary circuit interpretation, although it is conjectured to be related to dual random unitary circuit~\cite{Bertini:2018fbz,Piroli:2019umh}. We will therefore focus on ordinary membrane theory in the remainder of this paper. } 

% Notice that when the geometry \eqref{blackBraneInf} is BTZ black brane, $a(z)=1-z^2,\,b(z)=1$, giving $v_E=v_B= \mathcal{E}(v)=1$. 

% Let us consider an elementary example that will be used in later studies. Suppose we have the black brane geometry~\eqref{blackBraneInf}, with the entangling subregion taken to be half-spaces on both sides. The HRT surface stays on the $x=0$ plane. At late time, the HRT surface gets stuck at a special extremal slice $z=z_*$~\cite{Hartman:2013qma} (Figure~\ref{fig:DispHalfSpaceAdSSch} left),  and the entanglement entropy grows linearly with slope $2s_\text{th} \vol(\p A)\ v_E$. From the entanglement membrane point of view, we are minimising over all possible membranes connecting the two boundary points $(t,x)=(\pm T,0)$ (Figure~\ref{fig:DispHalfSpaceAdSSch} right). The minimal membrane is easily seen to be a verticle line with $v=0$. The membrane tension function \eqref{AreaFunct2} is therefore $\mathcal{E}(0)=v_E$. This vertical membrane can be obtained from the HRT surface by $\emph{projecting}$ the latter to the two boundaries along constant infalling time $u$, and $\emph{gluing}$ the two resulting projections at $u=0$. 

\begin{figure}[htbp]
\centering
\includegraphics[width=.44\textwidth]{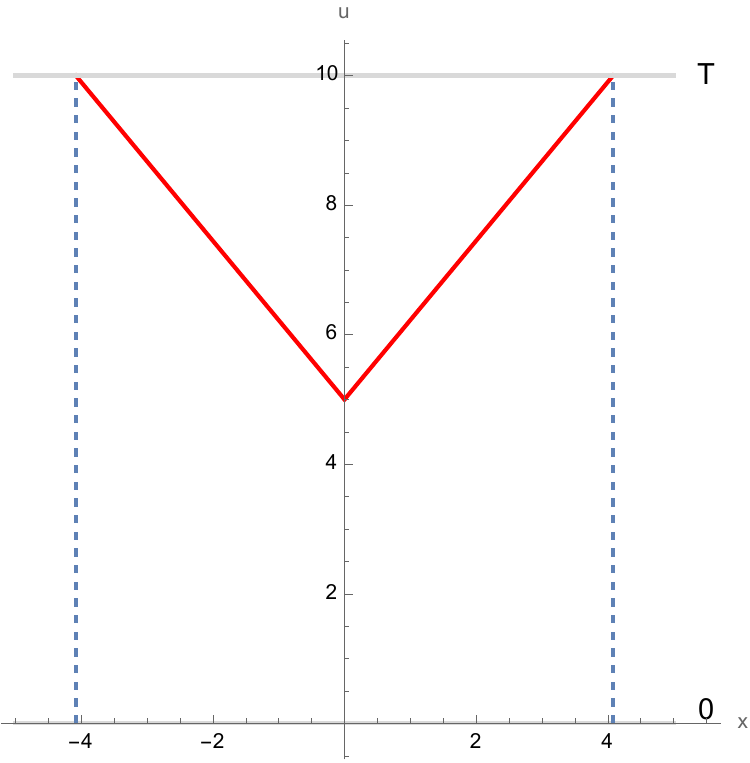}\hspace{5mm}
\includegraphics[width=.45\textwidth]{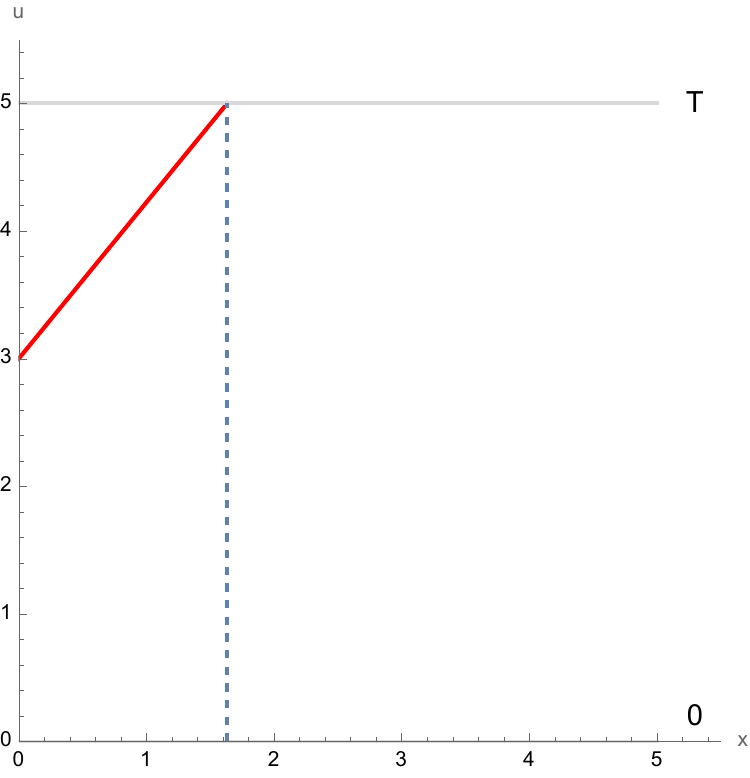}
\qquad
\caption{Two types of saturated entanglement membranes (red) in finite systems. In membrane theory, saturation of a finite system occurs when the red membranes with slope $\pm v_B$ compute smaller entanglement entropy compared to the dashed vertical membranes reaching the $t=0$ slice. These membranes with $\pm v_B$ slope can either form a cone (left) or exit the system via its boundary (right). In both cases, the saturation time is at $t=\frac{X}{v_E}$, where $X$ is the size of the entangling subregion. \label{fig:MembraneSat}}
\end{figure}

So far we have focused on semi-infinite systems. For finite systems, the entanglement entropy eventually $\emph{saturates}$. In membrane theory, saturation happens when the minimal membranes with $\pm v_B$ slope compute smaller entanglement entropy compared to those reaching the $t=0$ slice~\cite{Jonay:2018yei}. These minimal membranes can either form a $v_B$ $\emph{cone}$ (Figure~\ref{fig:MembraneSat} left), or exit the system via its boundaries (Figure~\ref{fig:MembraneSat} right). The saturation value of entanglement entropy they compute is $S=s_\text{th} \vol(\p A) X$, where $X\gg 1$ is the size of the entangling subregion in the boundary. Holographically, saturated HRT surfaces are static surfaces on constant-time slices outside the horizon; in the large entangling subregion limit, these static RT surfaces consist of $\emph{plateaux}$ close to the horizon, where the RT surfaces move transversely along the $x$ direction~\cite{Hubeny:2012ry,Mezei:2016zxg}. The $v_B$ cone saturated membranes shown in Figure~\ref{fig:MembraneSat} left correspond to projections of these static RT surfaces to the boundary along constant infalling time~\cite{Jiang:2024tdj} (see also~\cite{Mezei:2016wfz,Mezei:2016zxg}). To see this, we first notice that for the metrics~\eqref{blackBraneInf}, the infalling time along the $t=T$ slice is\footnote{The Schwarzschild time $t$ in the geometry \eqref{blackBraneInf} is defined as
\begin{equation}
t = u+\int_0^z\frac{dz'}{a(z')b(z')}
\end{equation}} 
\begin{align}
    u(z)&=T - \int_0^z \frac{dz'}{a(z')b(z')}= T-\frac{1}{a'(1)b(1)}\log (1-z)+...\label{uzStripExp}
\end{align}
where we have used $a(1)=0$. We parametrise the static RT surface in the black brane geometry~\eqref{blackBraneInf} as $x(z)$
\begin{align}
    x(z)=\int_0^{z}\frac{z'^{d-1}dz}{b(z')\sqrt{a(z') (z_0^{2 (d-1)}-z'^{2 (d-1)})}}=-\frac{\log (1-z)}{b(1)\, \sqrt{-2(d-1)a'(1)}}\label{xzStripExp}+...
\end{align}
where $z_0<1$ is the deepest point the RT surface can reach in the bulk outside the horizon. Combining \eqref{uzStripExp} and \eqref{xzStripExp} to eliminate $z$, we find 
\begin{align}
    x(u)=-\sqrt{-\frac{a'(1)}{2(d-1)}}u+...=-v_B u+...\label{ButterflyCone}
\end{align}
Indicating that the projections are lines with slope $v_B$. These conclusions remain true for arbitrary shapes of entangling subregions, as was implied implicitly in earlier results~\cite{Mezei:2016zxg,Mezei:2016wfz}. In 2d CFT, the above analysis still holds with $v_B=1$~\cite{Jiang:2024tdj}. Notice that in obtaining the $v_B$ slope~\eqref{ButterflyCone} we used $\emph{only}$ the fact that the RT surfaces get close to the horizon in the large entangling subregion limit. 

We emphasise that entanglement membranes from holography as depicted in e.g. Figure~\ref{fig:DispHalfSpaceAdSSch} and~\ref{fig:MembraneSat} only capture the shape of the HRT surfaces to $O(\Lambda)$, as the projection~\eqref{projectionz} itself is derived in the scaling limit~\eqref{CoordScaling}. A comparison between projections of exact and scaling surfaces can be found in~\cite{Jiang:2024tdj}. 

Finally, there is a direct connection between entanglement membrane and tensor network~\cite{Jonay:2018yei}: the minimal membrane shown in Figure~\ref{fig:MembraneCartoon} can be regarded as a coarse-grained cut through the tensor network representing the time-evolving state of interest, in which case the tension $\mathcal{E}(v)$ approximates the density of legs cut by the membrane.

\section{Set-up of the Double Holography Model}\label{DoubleHoloSetup}
\begin{figure}[htbp]
\centering
\includegraphics[width=.55\textwidth]{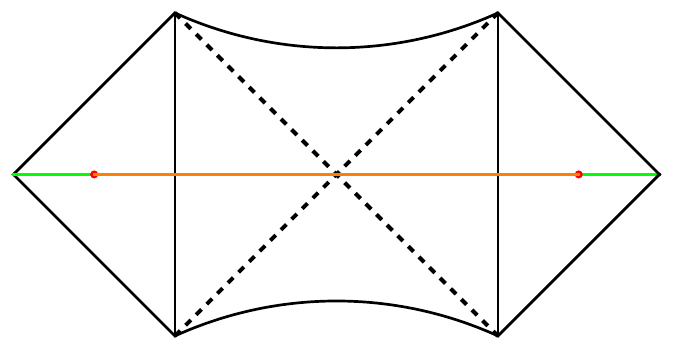}
\includegraphics[width=.4\textwidth]{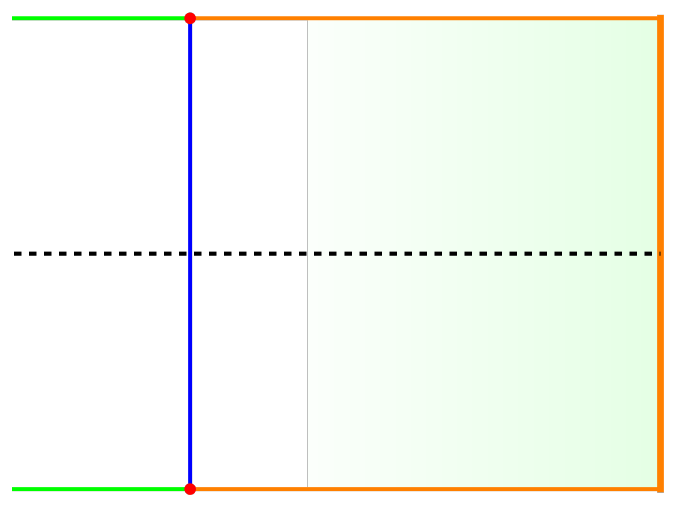}
\qquad
\caption{$\emph{Left:}$ We compute the Hawking radiation collected in the reservoir $(-\infty,-b]_{L}\cup [b,+\infty)_R$, shown as green. The complementary gravitating region $[-b,\infty)_L \cup (-\infty,b]_R$ is depicted in orange. The two red dots are the boundary points of the reservoir $x=\pm b$. $\emph{Right:}$ The double holography realisation of the left system, on the $t=0$ slice. The reservoir and gravitating region are plotted as green and orange, respectively, to match the left Penrose diagram. The Planck brane is plotted as thick in the right (in this cartoon we did not plot the angle $\theta$). The dashed line in the middle is the AdS$_{d+1}$ black hole horizon. In cases with $\theta<\frac{\pi}{2}$~\cite{Almheiri:2019psy}, we use green to denote the backreaction of the brane to the $(d+1)$ geometry, with darker colour corresponding to a stronger effect. As $b$ is large, the Hartman-Maldacena surface (blue) is in regions where the geometry is asymptotically AdS$_{d+1}$-Schwarzschild. \label{fig:EarlyTime}}
\end{figure}

As is reviewed in section~\ref{MembraneReview}, membrane theory derived from holography becomes degenerate in $d=2$ BTZ black hole. We will therefore focus on the $d>2$ cases~\cite{Almheiri:2019psy} in the remainder of this paper.\footnote{One can introduce a relevant deformation in the dual 2d CFT to restore the non-degenerate membrane tension~\cite{Jiang:2024tdj}. However, entanglement dynamics in the resulting conformal perturbation theory are similar to the $d>2$ cases. See~\cite{Caceres:2021fuw} for double-holographic models with relevant deformations to the bath. }  We consider an eternal two-sided asymptotically AdS$_d$ ($d>2$) black hole, coupled to two flat $d$-dimensional baths on each side. The AdS$_d$ and Mink$_d$ region are $[0,\infty)_L \cup (-\infty,0]_R$ and $(-\infty,0]_{L}\cup [0,+\infty)_R$,  respectively, where $L$ and $R$ denote left and right side. See Figure~\ref{fig:EarlyTime}. We take the $d$-dimensional bulk matter to be CFT$_d$, and impose transparent boundary conditions at $x=0$ on each side.  Field-theoretically, we can regard the two-sided asymptotically AdS$_d$ black hole as dual to two copies of maximally entangled CFT$_{d-1}$s defined on the conformal defects at $x=0$ on each side. 

We are interested in evaluating the Hawking radiation emitted from the black hole, collected in the reservoir $\mathfrak{R}=(-\infty,-b]_{L}\cup [b,+\infty)_R$ ($b>0$) in the non-gravitating bath region. The complementary region $[-b,\infty)_L \cup (-\infty,b]_R$, which is referred to as the gravitating system, contains the entire asymptotically AdS$_d$ black hole plus a part of the flat space bath region. See Figure~\ref{fig:EarlyTime}. The entropy of Hawking radiation collected in the reservoir $\mathfrak{R}$ is computed by the generalised entropy formula~\cite{Faulkner:2013ana}
\begin{align}
    S_{\rm{gen}}(\mathfrak{R})=\frac{A(\partial I)}{4G_N^{(d)}}+S_{{\rm bulk}}(\mathfrak{R}\cup I)\label{SGenBrane}
\end{align}
where $G_N^{(d)}$ denotes the Newton gravitational constant in $d$-dimensions, $I$ stands for the islands, and $S_{{\rm bulk}}$ is the entropy of bulk quantum fields in $d$-dimensions.\footnote{This "bulk" is not to be confused with the $(d+1)$-dimensional bulk in the double-holography model.} Nevertheless, the holographic derivation of membrane theory reviewed in section~\ref{MembraneReview} applies to HRT surfaces without quantum corrections. Moreover, apart from the special case of 2d CFT~\cite{Almheiri:2019psf,Almheiri:2019yqk}, $S_{{\rm bulk}}$ is difficult to compute directly. We are thus led to consider the special case where the matter CFT$_d$ has an AdS$_{d+1}$ gravity dual.\footnote{In $d>2$, the EE can also receive contributions from fluctuating gravitons. In the holographic matter limit, however, the large degrees of freedom make the graviton contribution subleading. } In such a case, the entropy of the reservoir $\mathfrak{R}$ can be computed double-holographically in AdS$_{d+1}$, where the QES in \eqref{SGenBrane} in AdS$_d$ gets geometrised to ordinary HRT surfaces. In the double holography model, the two flat baths serve as the two conformal boundaries of AdS$_{d+1}$; the asymptotically AdS$_d$ black hole is regarded as living on a Planck brane in AdS$_{d+1}$ via the Karch-Randall-Sundrum mechanism. 

%The two flat baths serves as the two conformal boundaries of the AdS$_{d+1}$ spacetime, and the AdS$_d$ spacetime in Figure \ref{fig:EarlyTime} is a Planck brane embedded in it.

Now let us describe the $(d+1)$-dimensional bulk setup in detail. We consider Einstein gravity in $(d+1)$-dimensions with a negative cosmological constant. We introduce a codimension-1 hypersurface $\mathcal{P}$ that serves as a $\emph{boundary}$ component of the $(d+1)$-dimensional spacetime. In AdS/BCFT~\cite{Takayanagi:2011zk,Fujita:2011fp}, this hypersurface $\mathcal{P}$ is identified as the Planck brane. The action is given by 
\begin{align}
    I=\frac{1}{16\pi G_N^{(d+1)}}\int d^{d+1}x\sqrt{-g}\Big(R+\frac{d(d-1)}{L^2}\Big)+\frac{1}{8\pi G_N^{(d+1)}}\int_{\mathcal{P}} d^{d}x\sqrt{-h}(K-\mathcal{T})\label{Action}
\end{align}
where $g_{ab}$ denotes the $d+1$-dimensional bulk metric, $G_N^{(d+1)}$ is the Newton gravitational constant in $(d+1)$-dimensions, and $L$ stands for the radius of AdS$_{d+1}$. The boundary term in \eqref{Action} is not to be confused with the Gibbons-Hawking boundary on $\partial$AdS$_{d+1}$ (which we omitted to avoid confusion): $\mathcal{P}$ is the Planck brane with a constant tension $\mathcal{T}$,\footnote{Such brane with a constant tension cannot support JT gravity on it in $d=2$. See~\cite{Grimaldi:2022suv} for the double holographic model for JT gravity. } and $h_{ab}$ is the induced metric on $\mathcal{P}$. 

% \footnote{In the $(d+1)$-dimensional bulk we still have $G_N^{(d+1)}\to 0$, so that holographic entanglement entropy is computed by ordinary codimension-2 HRT surfaces $S=\frac{{\rm Area(HRT)}}{4G_N^{(d+1)}}$.} 

Varying the action \eqref{Action} with respect to the metric leads to Einstein's equation in the bulk. As for the boundary term in \eqref{Action}, a standard treatment is to impose Dirichlet boundary condition $\delta h^{ab}=0$. Here, however, we instead impose $\emph{Neumann}$ boundary conditions~\cite{Takayanagi:2011zk,Fujita:2011fp} 
\begin{align}
    K_{ab}-K h_{ab}+\mathcal{T} h_{ab}=0\label{NeumannBC}
\end{align}
where $K_{ab}$ is the extrinsic curvature on $\mathcal{P}$ and $K$ is its trace. The boundary condition \eqref{NeumannBC} is a class of Israel junction conditions \cite{Israel:1966rt}. It is this junction condition \eqref{NeumannBC} that allows $\mathcal{P}$ to be identified as a Planck brane. 

\begin{figure}[htbp]
\centering
\includegraphics[width=.6\textwidth]{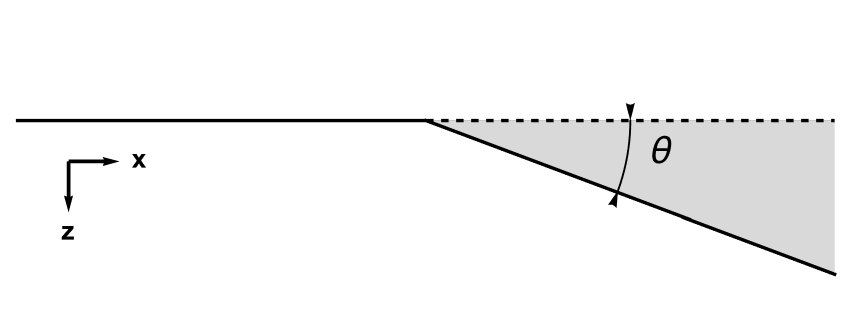}
\qquad
\caption{A Planck brane in vacuum AdS$_{d+1}$ with locus \eqref{braneLocus} at an angle $\theta$ to $\partial$AdS$_{d+1}$ on a constant time slice. The induced geometry on the brane is AdS$_d$. The shaded region is removed. \label{fig:AdSBCFTdemo}}
\end{figure}

The essentialities of AdS/BCFT correspondence~\cite{Takayanagi:2011zk,Fujita:2011fp} relies on the fact that in Poincare AdS$_{d+1}$, the hypersurface 
\begin{align}
    \mathcal{P}:z=x\tan\theta && \theta\in\big(0,\frac{\pi}{2}\big)\label{braneLocus}
\end{align}
at an angle $\theta$ to $\partial$AdS$_{d+1}$ is an AdS$_d$ foliation of AdS$_{d+1}$, see Figure \ref{fig:AdSBCFTdemo}. One can then compute the extrinsic curvature on $\mathcal{P}$ and use the junction condition \eqref{NeumannBC} to find the tension of the Planck brane $\mathcal{P}$, $\mathcal{T}=\frac{d-1}{L}\cos \theta$. The brane tension decreases with the angle $\theta$. In particular, when $\theta=\frac{\pi}{2}$, the brane ends on the AdS$_{d+1}$ boundary perpendicularly and becomes tensionless.  
% \begin{align}
%     ds^2=\frac{L^2}{z^2}(-dt^2+dz^2+dx^2+d\Vec{x}_{d-2}^2)
% \end{align}

\begin{figure}[htbp]
\centering
\includegraphics[width=.95\textwidth]{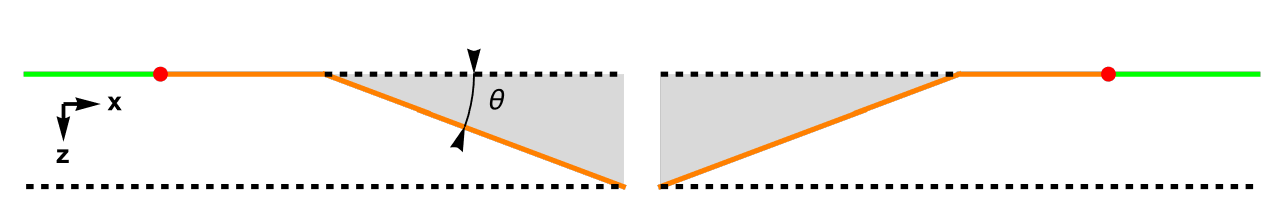}
\includegraphics[width=.94\textwidth]{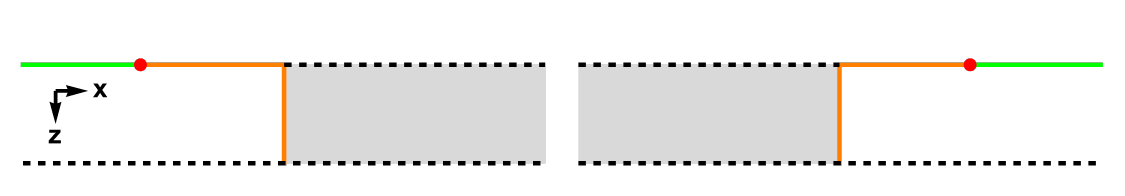}
\qquad
\caption{Eternal two-sided asymptotically AdS$_{d+1}$ black hole, each with a $d$-dimensional brane in the bulk at angle $\theta$ to the boundary. Here the lower dashed lines are the black hole horizon, and the shaded regions are removed. We again use green and orange to denote the reservoir and gravitating system from the $d$-dimensional point of view, respectively. Notice that we have plotted only the $\emph{exterior}$ geometry on a constant time slice. The brane on each side is expected to join in the interior, see i.e.~\cite{Chen:2020hmv}. The plots above and below denote the general $\theta<\frac{\pi}{2}$~\cite{Almheiri:2019psy} and the $\theta=\frac{\pi}{2}$ 'probe brane' case~\cite{Geng:2020qvw}, respectively. \label{fig:AdSBCFTBHdemo}}
\end{figure}

The above AdS/BCFT~\cite{Takayanagi:2011zk,Fujita:2011fp} construction is in the vacuum. In this work, nevertheless, we will be interested in the Hartle-Hawking state dual to an eternal two-sided asymptotically AdS$_{d+1}$ ($d>2$) black\ hole. Like in the vacuum case, we take the brane $\mathcal{P}$ location to be at $z=x\tan\theta$ \eqref{braneLocus} outside the horizon, and impose Neumann boundary condition \eqref{NeumannBC} on the brane $\mathcal{P}$, as is in~\cite{Almheiri:2019psy}. See Figure~\ref{fig:AdSBCFTBHdemo} above. In $d>2$, one would expect the Planck brane $\mathcal{P}$ to $\emph{backreact}$ the local spacetime surrounds it. Indeed, solving Einstein's equation with the boundary condition \eqref{NeumannBC} at $\mathcal{P}$ is nontrivial. In $d=4$, such metric is found numerically~\cite{Almheiri:2019psy} using the DeTurck trick~\cite{Headrick:2009pv}.\footnote{We expect the gravitational solutions in $d>2$ to be qualitatively similar.}  Once we back away from the brane, however, the $(d+1)$-dimensional bulk metric asymptotes that of the planar AdS$_{d+1}$-Schwarzschild black hole~\cite{Almheiri:2019psy}.
\begin{align}
    ds^2=\frac{1}{z^2}\Big[-f(z)du^2-2dudz+dx^2+d\Vec{y}_{d-2}^2\Big]\label{AdSSch}
\end{align}
where $f(z)=1-z^d$. This simplification will play a crucial role in our later studies. 

In general, the induced gravity on the brane $\mathcal{P}$ is not Einstein gravity, but contains higher curvature corrections~\cite{Chen:2020uac}. In our work, however, we will perform calculations in the AdS$_{d+1}$ bulk, therefore bypassing this subtlety. What is important to our calculation is that the $d+1$-dimensional black hole induces a $d$-dimensional black hole on the brane $\mathcal{P}$, whose horizon is the intersection surface between the horizon of the $d+1$-dimensional black hole and the $d$-dimensional brane~\cite{Almheiri:2019psy}.

To sum up, we have three equivalent ways of viewing our double holographic model, 
\begin{itemize}
    \item Bulk perspective: Asymptotically AdS$_{d+1}$ black hole with a $d$-dimensional brane $\mathcal{P}$ in the bulk. The $(d+1)$ geometry approaches that of planar AdS$_{d+1}$-Schwarzschild black hole away from the brane $\mathcal{P}$. 
    \item Brane perspective: Asymptotically AdS$_{d}$ black hole coupled to two thermal baths in flat $d$-dimensional spacetime. 
    \item Boundary perspective: 2 copies of $d$-dimensional CFTs in Minkowski coordinates, each coupled to a $d-1$-dimensional conformal defect at $x=0$. 
\end{itemize}
While the brane perspective is our starting point in Figure~\ref{fig:EarlyTime} left, in the remainder of this paper we will focus on the $(d+1)$-dimensional bulk perspective where Einstein gravity serves as a technical tool for calculating the generalised entropy~\eqref{SGenBrane} geometrically. 

A particularly simple limit of our double holographic model, known as the 'probe brane' case, is when $\theta=\frac{\pi}{2}$~\cite{Geng:2020qvw} (see Figure~\ref{fig:AdSBCFTBHdemo} below). In this case, the probe brane is at $x=0$. The extrinsic curvature $K_{ab}=0$ and the brane becomes $\emph{tensionless}$ $\mathcal{T}=0$ due to~\eqref{NeumannBC}. The probe brane does not backreact on the $(d+1)$-dimensional geometry: the metric is simply that of planar AdS$_{d+1}$-Schwarzschild black hole~\eqref{AdSSch}. While the triviality at $x=0$ seems to indicate that there is no brane at all, one can think of $x=0$ as an orbifold plane~\cite{Geng:2020qvw}: the $(d+1)$-dimensional geometry is obtained from planar AdS$_{d+1}$-Schwarzschild black hole~\eqref{AdSSch} by orbifolding $x\to -x$. Likewise, the $d$-dimensional field theory can be constructed from holographic CFT$_d$ by orbifolding $x\to -x$.\footnote{For specific dualities such as that between $\mathcal{N}=4$ SYM and type IIB superstring theory on AdS$_5\times$S$^{5}$~\cite{Maldacena:1997re}, this is the only case in which the dual CFT is explicitly known~\cite{Geng:2020qvw}.} The $d$-dimensional gravity living on the probe brane at $x=0$ is a non-standard theory of (massive) quantum gravity~\cite{Geng:2020qvw}. Nonetheless, our later analysis will not involve its detail, since we will calculate the entropy of radiation geometrically in the $(d+1)$-dimensional bulk, whose metric is~\eqref{AdSSch}.

\section{Entanglement membrane from double holography}\label{PageCurveMembrane}
Now, having turned the problem concerning quantum effects in $d$-dimensions to a geometric one in $(d+1)$-dimensions, we are ready to execute the projections of the HRT surface to $\partial$AdS$_{d+1}$ along constant infalling time as is in membrane theory~\cite{Mezei:2018jco}. For simplicity, we will first focus on probe brane, and then discuss the general cases when $\theta<\frac{\pi}{2}$.

\subsection{Probe brane case}\label{MembraneProbeBrane}

\begin{figure}[htbp]
\centering
\includegraphics[width=.37\textwidth]{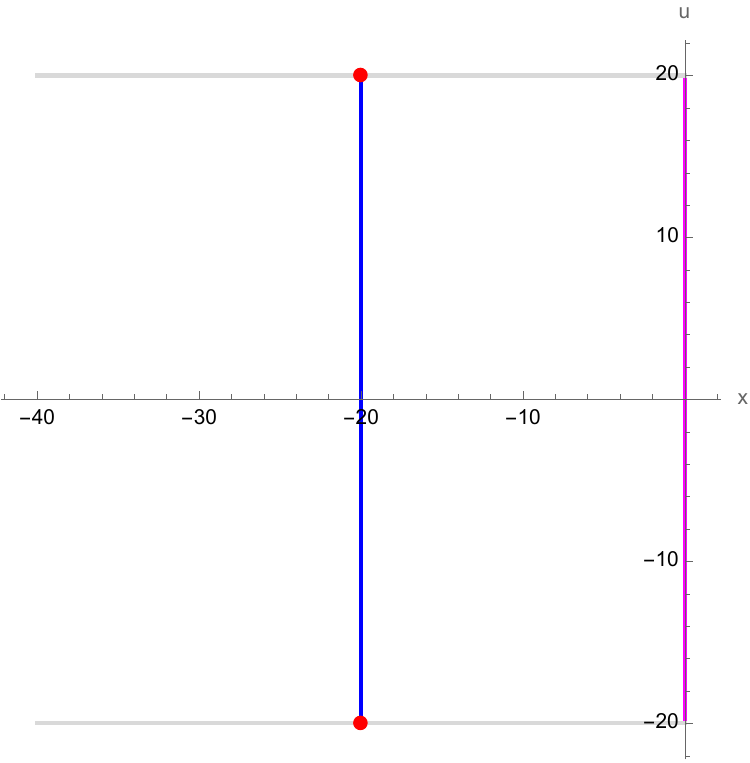}\hspace{5mm}
\includegraphics[width=.37\textwidth]{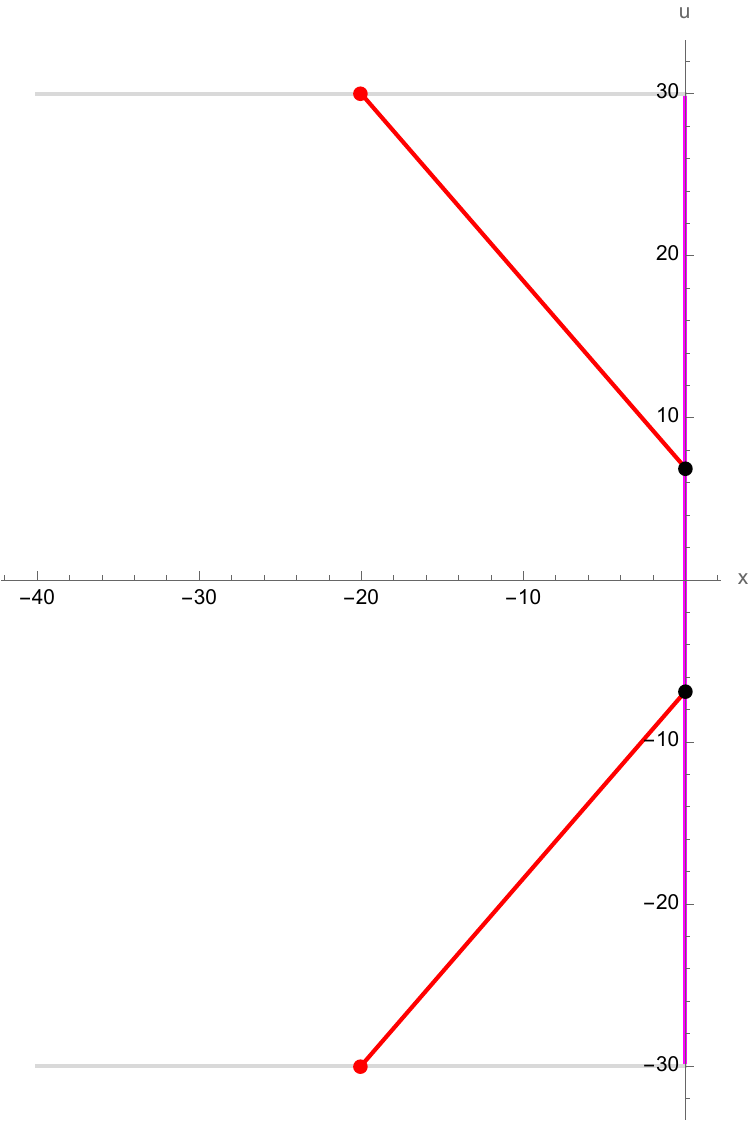}
\qquad
\caption{Projections of the HM surface (blue) and island surface (red) to $\partial$AdS$_{d+1}$ along constant infalling time, at $T=20$ (left) and $T=30$ (right), respectively. The probe brane is projected to the $x=0$ boundary (Magenta) on the right. In the right plot, the two black dots denote the exit positions of the $\pm v_B$ slope lines at the $x=0$ boundary. Here $d=3$ and $b=20$. \label{fig:MembraneDoubleHolo}}
\end{figure}

In the probe brane limit, the $(d+1)$-dimensional geometry is exactly that of planar AdS$_{d+1}$ black hole~\eqref{AdSSch}. The entropy of Hawking radiation collected in $\mathfrak{R}$ is computed double-holographically by the codimension-2 ordinary HRT surfaces in AdS$_{d+1}$ homologous to $\mathfrak{R}$, which goes through the interior of the two-sided AdS$_{d+1}$ black hole, see Figure~\ref{fig:EarlyTime} right. Due to the stretch of spacelike interior-$t$ direction~\cite{Hartman:2013qma}, the entanglement entropy grows linearly in time with slope $v_E$.  We will refer to this HRT surface as Hartman-Maldacena (HM) surface. From the AdS$_d$ brane + Mink$_d$ bath perspective, this corresponds to having no island in~\eqref{SGenBrane}, see Figure~\ref{fig:EarlyTime} right. We have seen in Figure~\ref{fig:DispHalfSpaceAdSSch} that the HM surface is described by membrane theory in the late-time, i.e. $t\to \Lambda t$ limit: its projection to $\partial$AdS$_{d+1}$ is simply a vertical line. This $v=0$ membrane computes the entanglement entropy
\begin{align}
    S(\mathfrak{R})=2s_\text{th} \vol(\p A)v_E t\label{HMEE}
\end{align}
and is plotted as blue in Figure~\ref{fig:MembraneDoubleHolo}. Here $v_E=\frac{\big(\frac{d-2}{d}\big)^{\frac{d-2}{2d}}}{\big(\frac{2(d-1)}{d}\big)^{\frac{d-1}{d}}}$ and $s_\text{th}=\frac{1}{4G_N^{(d+1)}}$ are the entanglement velocity and thermal entropy of planar AdS$_{d+1}$-Schwarzschild black hole, respectively.

\begin{figure}[htbp]
\centering
\includegraphics[width=.55\textwidth]{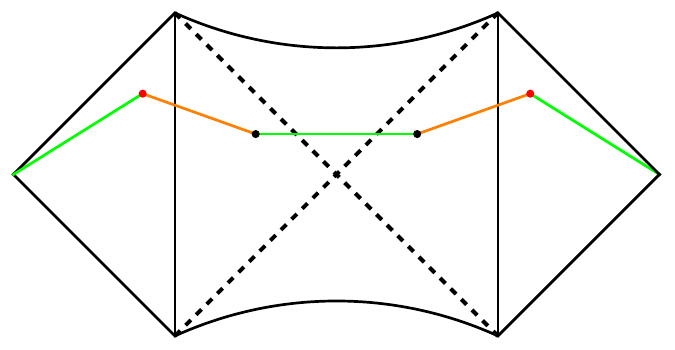}
\includegraphics[width=.4\textwidth]{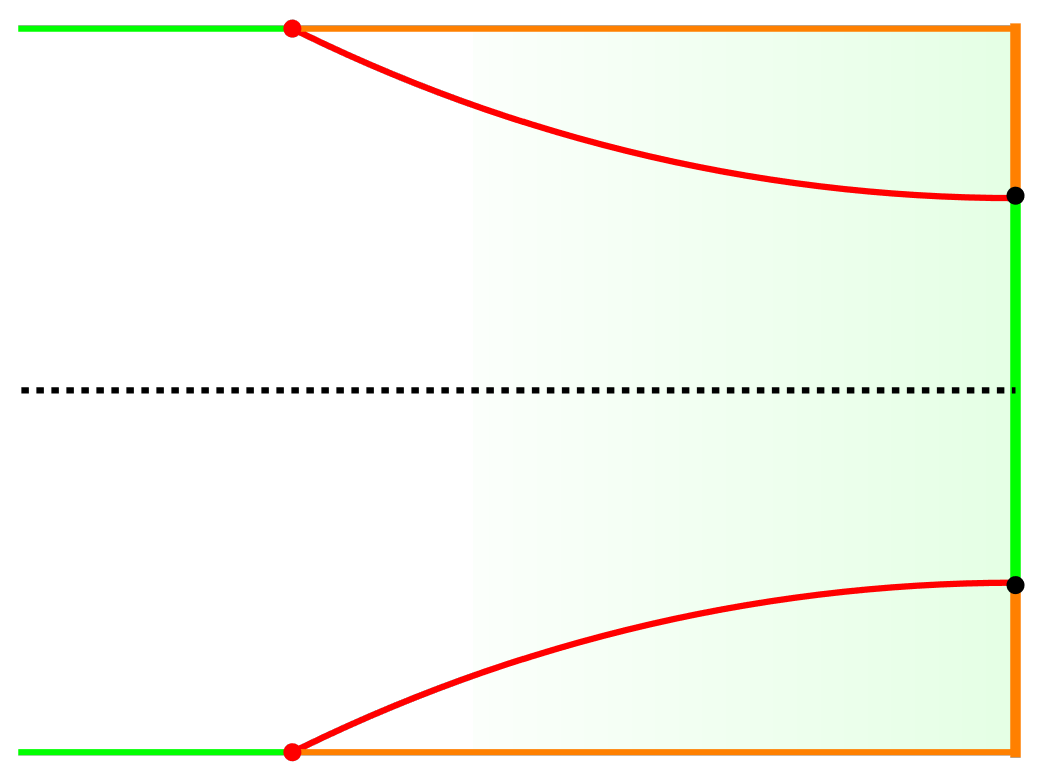}
\qquad
\caption{$\emph{Left:}$ Emergence of the island (green interval in the middle) after the Page time. To compute the entropy of Hawking radiation collected in the reservoir $(-\infty,-b]_{L}\cup [b,+\infty)_R$, one must also include contributions from the island $[a,+\infty)_{L}\cup (-\infty,-a]_R$ in the gravitating region disconnected to the bath. The EW of the radiation and gravitating system are the causal domain of dependence of the green and orange regions, respectively. The two black dots $x=\pm a$ denote the location of the QES. $\emph{Right:}$ The double holography realisation of the left system, on a constant-$t$ slice. The island surfaces (red) end on the Planck brane. Notice that the disconnected island region in the left Penrose diagram becomes connected in the right cartoon through the extra dimension. In the $\theta<\frac{\pi}{2}$ case~\cite{Almheiri:2019psy}, when $b$ is large, the portion of island surfaces receiving corrections from the brane backreaction is subleading. \label{fig:LateTime}}
\end{figure}

After the Page time $t_P$, however, there will be a second class of HRT surfaces that have a smaller area and therefore dominate. These are $\emph{static}$ RT surfaces ending on the Planck brane $\mathcal{P}$~\cite{Almheiri:2019psy}, see Figure~\ref{fig:LateTime} right. The entanglement entropy they compute does not grow in time as they avoid the interior growth of the AdS$_{d+1}$ black hole. From the AdS$_d$ braneworld gravity point of view, there is an island emerging at the place where these static RT surfaces hit the brane outside the $d$-dimensional horizon, as is depicted in Figure~\ref{fig:LateTime} left. We will therefore denote these static RT surfaces as island surfaces. Thus, in double holography, the quantum extremal island rule~\cite{Penington:2019npb,Almheiri:2019psf,Almheiri:2019hni} is geometrised to the competition between ordinary HRT surfaces in the $(d+1)$-dimensional bulk. In particular, the island surfaces geometrise $\emph{both}$ terms in~\eqref{SGenBrane}~\cite{Almheiri:2019psy}.\footnote{Notice that this geometrisation is different compared to that in~\cite{Almheiri:2019hni}, as the latter studied JT gravity whose brane profile is not~\eqref{Action} with a constant tension.} 

\begin{figure}[htbp]
\centering
\includegraphics[width=.6\textwidth]{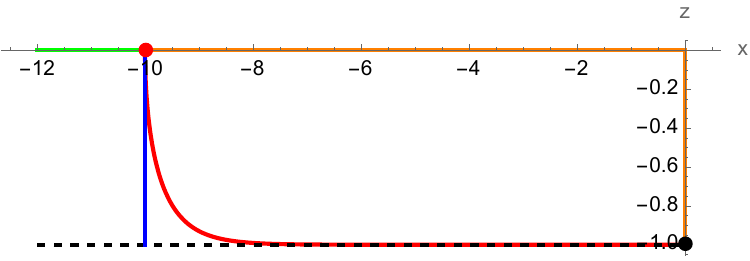}
\qquad
\caption{An illustration of the HM surface (blue) and island surface (red) in the probe brane limit, on a constant time slice $t=T$ outside the horizon, in one side of the geometry shown in Figure~\ref{fig:AdSBCFTBHdemo} below. The probe brane at $x=0$ can be regarded as an orbifold plane since we excise the $x>0$ region. The island surface is characterised by the integral expression~\eqref{xzStripExp}, orbifolded by $x\to -x$. Consequently, it can be regarded as half of an ordinary RT surface anchored on $\partial$AdS$_{d+1}$ at $x=\pm b$ (the point at $x=-b$ is plotted as a red dot), ending on the probe brane $x=0$ perpendicularly at the deepest point $z_0=a$ it reaches into the bulk (black dot). As $b$ is large, the island surface contains a large plateau where it skims through the horizon (dashed black line). \label{fig:ProbeBraneIslandSurf}}
\end{figure}

As is reviewed in section~\ref{MembraneReview}, membrane theory only applies in the scaling limit~\eqref{CoordScaling}. As for the static island surfaces, one needs to scale $b$ large:\footnote{This limit is consistent with the fact that for the Page curve to appear, i.e. for the island surfaces to have a greater area than the $t=0$ HM surface, $b$ needs to be larger than a threshold value~\cite{Geng:2020qvw}. In our scale, however, this threshold value is $O(1)$~\cite{Geng:2020qvw}. So we will take $b$ even larger. }  
\begin{align}
    b\to \Lambda b
\end{align}
As the $(d+1)$-dimensional geometry is that of planar AdS$_{d+1}$-Schwarzschild black hole, the island surfaces are characterised by~\eqref{xzStripExp}, containing large plateaux skimming through the horizon (see Figure~\ref{fig:ProbeBraneIslandSurf}). Therefore, projections of these surfaces to the boundary are lines of $\pm v_B$ slope~\eqref{ButterflyCone}~\cite{Jiang:2024tdj} and are plotted as red in Figure~\ref{fig:MembraneDoubleHolo}. There is, however, one important difference between the island surface and ordinary static RT surfaces: the former ends on the probe brane $x=0$ instead of returning to the AdS$_{d+1}$ boundary. Due to the $x\to -x$ symmetry, the island surface can be regarded as half of an ordinary RT surface~\eqref{xzStripExp} with width $2b$, ending on the probe brane $x=0$ perpendicularly. See Figure~\ref{fig:ProbeBraneIslandSurf}. Notice that the probe brane is projected to the surface $x=0$ on $\partial$AdS$_{d+1}$, which serves as the boundary of the $d$-dimensional field theory. Therefore, one can understand the membranes with slope $v_B$ ending on $x=0$ as exiting the system via this boundary, see Figure~\ref{fig:MembraneDoubleHolo}. This interpretation precisely realise the saturation mechanism in membrane theory depicted in Figure~\ref{fig:MembraneSat} right~\cite{Jonay:2018yei}. In the large $b$ limit, the entanglement entropy these two island surfaces jointly contribute is given by~\cite{Jonay:2018yei,Jiang:2024tdj}
\begin{align}
    S(\mathfrak{R})=2S_\text{Sat}=2s_\text{th}\vol(\p A)b\label{IslandEE}
\end{align}
which is time-independent and can be interpreted as the coarse-grained entropy of the gravitating system.\footnote{We avoid using the terminology $2S_\text{BH}$ as in~\cite{Almheiri:2019yqk,Almheiri:2019psy}, since the gravitating system with large $b$ contains not only the two-sided AdS$_d$ black hole but also a large region from the flat bath, see Figure~\ref{fig:EarlyTime}.} 

Putting \eqref{HMEE} and \eqref{IslandEE} together, we have the Page curve in the scaling limit~\eqref{scaling} computed from the membrane configuration in Figure~\ref{fig:MembraneDoubleHolo}
\begin{equation}
\label{PageCurve}
    S(\mathfrak{R})=2s_\text{th} \vol(\p A)
    \left\{
    \begin{aligned}
        &v_E t,&& t\leq t_{P}\\
        &b&& t\geq t_{P}
    \end{aligned}
    \right.
\end{equation}
where the Page time is 
\begin{align}
    t_P=\frac{b}{v_E}
\end{align}
Notice that we toroidally compact the traverse dimensions to get IR-finite entanglement entropy~\cite{Almheiri:2019psy}. 

The simplicity of the $(d+1)$-dimensional geometry allows us to address the location of the QES $z=a$ on the $d$-dimensional probe brane at $x=0$. By $x\to -x$ symmetry, the endpoint of the island surface on the probe brane is also the deepest point $z_0$ the island surface reaches in the $(d+1)$-dimensional bulk, see Figure~\ref{fig:ProbeBraneIslandSurf}. As the island surfaces get very close to the horizon when $b$ is large, one would expect the QES to be also very close to the horizon in the $d$-dimensional braneworld gravity. Indeed, a near-horizon expansion~\eqref{xzStripExp} of the island surface in the scaling limit establishes the following relation between $a$ and $b$  
\begin{align}
    b\approx -\frac{\log (1-a)}{\sqrt{-2(d-1)f'(1)}} && \Rightarrow && a\approx 1-e^{-\frac{d}{v_B}b}\label{QESLoc}
\end{align}
As $b$ increases, $a$ gets exponentially closer to the horizon, see Figure~\ref{fig:ProbeBraneIslandSurf}.

The Page curve~\eqref{PageCurve} is similar to that of eternal two-sided AdS$_{d+1}$-Schwarzschild black hole~\cite{Almheiri:2019yqk,Almheiri:2019psy}, except the gravitating system with large $b$ contains a region of considerable size from the flat bath in addition to the two-sided AdS$_d$ black hole. Therefore, by taking the large $b$ limit so as to apply membrane theory to island surfaces, we have changed the problem of the information paradox for an eternal black hole~\cite{Almheiri:2019yqk} to a similar information paradox concerning a finite system containing an eternal black hole. This information paradox results from the fact that any finite system consisting of a black hole can store only a finite number of degrees of freedom in it. As the quantum extremal island rule~\cite{Penington:2019npb,Almheiri:2019psf,Almheiri:2019hni} describes generic entanglement behaviours between quantum matters in one region and in another region with gravity, we can apply this rule to address our version of the information paradox the same way as in~\cite{Almheiri:2019yqk,Almheiri:2019psy} for the information paradox of the eternal black hole. Indeed, there are no threshold values of $b$ distinguishing the Page curve~\eqref{PageCurve} and that for an eternal black hole~\cite{Almheiri:2019yqk,Almheiri:2019psy}. When we take $b$ to be smaller,\footnote{By taking $b$ smaller, we still need to ensure it to be larger than the threshold value for the Page curve to occur~\cite{Geng:2020qvw}.} the entanglement entropy the island surfaces compute are closer to the coarse-grained entropy of the eternal two-sided AdS$_d$ black hole $2S_\text{BH}$. Nonetheless, as the island surfaces are small in this case, membrane theory does not apply anymore.

Built on earlier intuitions~\cite{Hartman:2013qma}, it was argued in~\cite{Almheiri:2019hni} that the configuration consisting of quantum systems in some regions and gravity in some other regions can be regarded as a tensor network preparing the state. Namely, one can understand the entanglement between these two regions as links of that tensor network. As entanglement membrane has a natural tensor network interpretation~\cite{Jonay:2018yei,Mezei:2018jco}, our double holographic membrane theory depicted in Figure~\ref{fig:MembraneDoubleHolo} is a concrete realisation of such a tensor network. 

\subsection{Comments on the $\theta<\frac{\pi}{2}$ cases}\label{backreactedCase}
When $\theta< \frac{\pi}{2}$, the $(d+1)$-dimensional bulk geometry is much more complicated than that of planar AdS$_{d+1}$-Schwarzschild black hole~\eqref{AdSSch} and does not admit an analytic expression, see~\cite{Almheiri:2019psy} for a numerical solution in $d=4$. However, as we will see, in the scaling limit, 
\begin{align}
    t\to \Lambda t && b\to \Lambda b \label{scaling}
\end{align}
the resulting membrane theory is qualitatively the $\emph{same}$ as that in the probe brane case shown in Figure~\ref{fig:MembraneDoubleHolo}. This is because the $(d+1)$-dimensional geometry asymptotes that of AdS$_{d+1}$-Schwarzschild black hole~\eqref{AdSSch} as one moves away from the brane~\cite{Almheiri:2019psy}. These conclusions apply provided that in addition to considering $b \gg 1$ as in ordinary membrane theory~\cite{Mezei:2018jco},\footnote{What this scaling really means is $\Lambda\gg \beta$, where $\beta$ is the inverse temperature of the black hole. As we have set $z_+=1$, $\beta\in O(1)$. } we also take $b$ to be much larger than the distance scale, $\Gamma$, over which the backreaction of the brane has a significant effect on the geometry. 

Let us first consider the HM surface before the Page time. Although the Planck brane breaks translational invariance in $x$ and backreacts the local geometry, if we take the large $b$ limit~\eqref{scaling} with $b \gg \Gamma$ then the HM surface will be located in regions where the $(d+1)$-dimensional bulk geometry is well approximated by the AdS$_{d+1}$-Schwarzschild black hole~\eqref{AdSSch}, see Figure~\ref{fig:EarlyTime} right. Therefore, to $O(\Lambda)$ the HM surface is the same as that in the probe brane limit. The EE it computes is also given by~\eqref{HMEE}. 

The island surface requires more care. Before proceeding, a subtlety to notice is that although the CFT$_d$ is defined on the half line $x<0$, when $\theta<\frac{\pi}{2}$ HRT surfaces can reach the $x>0$ regions in the $(d+1)$-dimensional bulk, see Figure~\ref{fig:AdSBCFTBHdemo}. Therefore, when projecting the island surface to $\partial$AdS$_{d+1}$, one needs to $\emph{extend}$ the $x=0$ boundary to some $x=x_b$ in the $x>0$ region. While the precise expression of $x_b$ requires knowledge of the full numerical metric, one can argue that it is of $O(1)$ value by noticing that the allowed bulk region scales as $z_+\in O(1)$ when $\theta$ is not taken to be small.\footnote{There are subtleties of divergence with island surfaces when $\theta\to 0$~\cite{Geng:2021mic}, so we will $\emph{not}$ scale $\theta\to\frac{1}{\Lambda}\theta$. } Thus, this extended region is $\emph{subleading}$ in membrane theory which devotes to $O(\Lambda)$ scale. In what follows, we will denote the extended boundary $x_b>0$ only for conceptual purpose while performing calculations at $O(\Lambda)$ as in membrane theory. 

As is in the probe brane case, taking the large $b$ limit~\eqref{scaling} allows the island surfaces to be large and therefore described by membrane theory, as is reviewed in section~\ref{MembraneReview}. Moreover, for $b \gg \Gamma$ the majority of the island surfaces lie away from the Planck brane in regions where the $(d+1)$-dimensional geometry can be treated as AdS$_{d+1}$-Schwarzschild black brane. As such the the portion of island surfaces receiving corrections from the brane backreaction is subleading.

Now let us focus on the projections of the island surfaces. As the majority of the island surfaces lie in the bulk regions far away from the Planck brane in the scaling limit~\eqref{scaling}, the near-horizon expansion~\eqref{ButterflyCone} also applies to them. Therefore, just as in the probe brane case, the projections of island surfaces to $\partial$AdS$_{d+1}$ along constant $u$ are lines with $\pm v_B$ slope. Notice that these membranes only reflect the geometry of the island surface to $O(\Lambda)$. Near the extended boundary $x_b$ where the membranes exit, one would expect the membrane shape to be affected by the brane backreaction. Nevertheless, these are $O(1)$ effects that are subleading in large $b$ limit. To investigate them, one needs to invoke the full numerical $(d+1)$-dimensional metric in~\cite{Almheiri:2019psy}. We thus conclude that for the purpose of membrane theory, i.e. to $O(\Lambda)$, the island surface is the same as that in the probe brane limit as well. The EE they compute is consequently also~\eqref{IslandEE}.\footnote{In~\cite{Almheiri:2019psy}, it was pointed out that for the island surface, there is a continuous family of extremal surfaces and there is a unique one among them with the smallest area. We believe that in the large $b$ limit~\eqref{scaling}, the difference among this family of extremal surfaces is an $O(1)$ effect: all island surfaces within this family are dominated by plateaux skimming through the horizons. Therefore, the entanglement entropy they compute are all given by~\eqref{IslandEE} to $O(\Lambda)$.}  

To sum up, the membrane theory computing the Page curve~\eqref{PageCurve} obtained from the probe brane limit depicted in Figure~\ref{fig:MembraneDoubleHolo} also applies to the $\theta<\frac{\pi}{2}$ cases. This is because the brane angle and backreaction are only $\emph{local}$ effects near the Planck brane. By taking the large $b$ limit~\eqref{scaling}, and further $b \gg \Gamma$, we have scaled out their influences on the HRT surfaces of interest. Thus, the membrane theory derived from the $\theta=\frac{\pi}{2}$ probe brane case in Figure~\ref{fig:MembraneDoubleHolo} can be regarded as a toy model for the generic $\theta<\frac{\pi}{2}$ cases to $O(\Lambda)$. The location of the QES~\eqref{QESLoc}, however, does not apply anymore due to the ignorance of the backreacted $(d+1)$-dimensional bulk metric near the brane. 

\subsection{Relation to the Blake-Thompson Model}\label{BTreview}
In this subsection, we compare the Blake-Thompson model~\cite{Blake:2023nrn} in chaotic many-body system to our double holographic model in Figure~\ref{fig:MembraneDoubleHolo}. We will consider the probe brane case in subsection~\ref{MembraneProbeBrane}, as the general $\theta<\frac{\pi}{2}$ cases discussed in subsection~\ref{backreactedCase} are qualitatively the same. 

\begin{figure}[htbp]
\centering
\includegraphics[width=.6\textwidth]{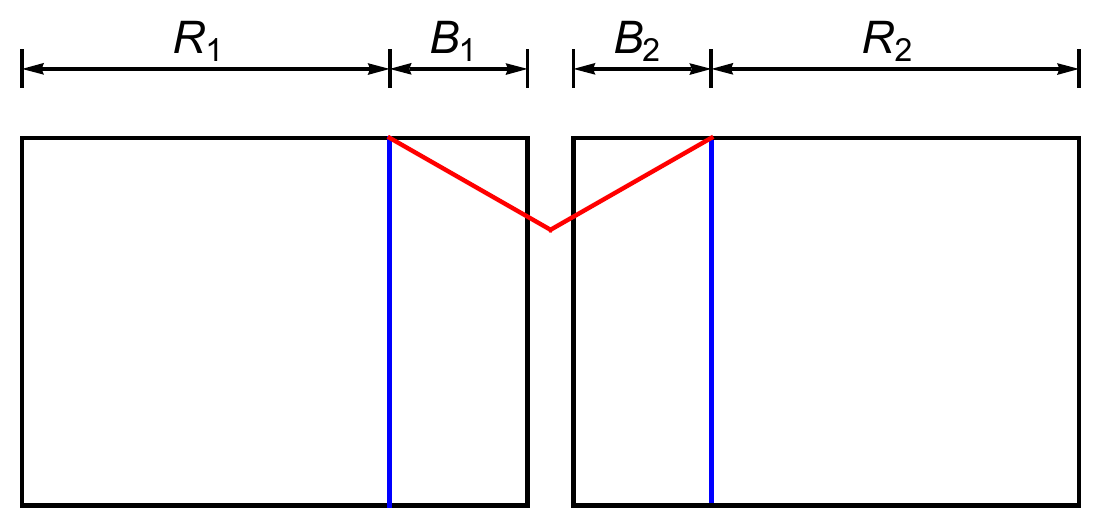}
\qquad
\caption{The Blake-Thompson model of eternal two-sided black holes ($B_1$ and $B_2$) each coupled to a bath ($R_1$ and $R_2$). \label{fig:BTModel}}
\end{figure}

The Blake-Thompson model consists of two maximally entangled chaotic many-body systems $B=B_1\cup B_2$ in reminiscent of an eternal two-sided black hole, each with size $\alpha$ and equilibrium entropy density~$\Tilde{s}_\text{th}$. Each $B_i$ is coupled to a bath $R_i$ ($i=1,2$), which is another chaotic many-body system with a different equilibrium entropy density~$s_\text{th}$,\footnote{The work~\cite{Blake:2023nrn} took the bath to be in a product state. To make a closer analogy to the gravity model, however, we will consider the bath to be instead in a thermofield double state. Results considering the entanglement of $R=R_1\cup R_2$ remain unchanged. } see Figure~\ref{fig:BTModel}. The quantity of interest is the entanglement entropy of the radiation region $S(R)$, where $R=R_1\cup R_2$. There are two groups of relevant membranes: two vertical membranes lying infinitesimally inside $R_1$ and $R_2$ (coloured blue in Figure~\ref{fig:BTModel}), computing an entropy $S(R)=2s_\text{th}\mathcal{E}(0)t=2s_\text{th}v_E t$; and two lines of slope $\pm v_B$ inside the black hole region (coloured red in Figure~\ref{fig:BTModel}), with an entropic contribution $S(R)=2S_{BH}=2\Tilde{s}_\text{th}\alpha$. Minimizing the entanglement entropy these two membrane configurations compute, we have the Page curve

\begin{equation}
\label{BTEE}
    S(R,t)=
    \left\{
    \begin{aligned}
        &2s_\text{th}v_E t,&& t\leq t_{P}\\
        &2S_{BH},&& t\geq t_{P}
    \end{aligned}
    \right.
\end{equation}
where the Page time is $t_P=\frac{S_{BH}}{s_\text{th}v_E}$. 

The Blake-Thompson model shown in Figure~\ref{fig:BTModel} is closely related to our double holographic entanglement membrane model depicted in Figure~\ref{fig:MembraneDoubleHolo}. To elucidate this connection, one can start by noticing that in the double holography model~\eqref{Action}, from the $(d+1)$-dimensional bulk point of view the probe brane serves as a $\emph{boundary}$ for the island surfaces to end on~\cite{Almheiri:2019psy}; when projected to $\partial$AdS$_{d+1}$, this boundary allows the membrane with $\pm v_B$ slope to $\emph{exit}$~\cite{Jonay:2018yei}. Therefore, in a membrane theory projected from the double holography setup~\eqref{Action} with a $\mathcal{T}=0$ probe brane, there are no 'black hole' degrees of freedom, only a homogeneous region with a boundary at $x=0$. To modify the Blake-Thompson model to our model, one can either let $R$ and $B$ be the same chaotic many-body system by setting $\Tilde{s}_\text{th}=s_\text{th}$, or discard $B$ and back off the reservoir a distance $b$ from the $x=0$ boundary. Afterward, one can glue the two-sided setup along the $u=0$ slice, as in Figure~\ref{fig:DispHalfSpaceAdSSch}. The resulting membrane theory is depicted in Figure~\ref{fig:MembraneDoubleHolo}.

\section{Conclusion and Outlook}
In this paper, we established an entanglement membrane model derived from a double holographic construction. By applying the quantum extremal island rule, we computed the Page curve for a system containing an eternal two-sided black hole in our model, thereby partially resolving the information paradox arising from this system. As our entanglement membrane model is obtained from a direct projection of the double holographic HRT surfaces to the boundary along constant infalling time, we established a $\emph{quantitative\ equivalence}$ between a chaotic quantum many-body system and a semi-classical gravity calculation of the Page curve. In our model, a particular novelty is the interpretation of the saturated membranes as two lines with $\pm v_B$ slope exiting the system through its boundary~\cite{Jonay:2018yei} originating from the Planck brane. 

% Additionally, we established a realisation of the Hayden-Preskill protocol in our double-holographic entanglement membrane theory by keeping track of the projection of the bulk EW. 

% Membrane theory of entanglement dynamics~\cite{Jonay:2018yei,Mezei:2018jco} entails taking the scaling limit, and captures the geometric shapes of the HRT surfaces to $O(\Lambda)$. In our model, taking $b\to\Lambda b$~\eqref{scaling} not only achieves this requirement, but also scales out the effects of the brane backreaction. 

We established an extremely simple double holographic membrane theory from the 'probe brane' case~\cite{Geng:2020qvw}, and argued that for generic $\theta<\frac{\pi}{2}$~\cite{Almheiri:2019psy} the membrane theory is the same to $O(\Lambda)$, where $\Lambda\gg 1$ is also much larger than the backreaction scale $\Gamma$. From this point of view, the backreaction of the Planck brane can be regarded as a detail of the specific gravitational solution~\cite{Almheiri:2019psy} instead of a universal feature of entanglement dynamics in generic chaotic systems. While we managed to extract these universal aspects in our double holographic model in the $\theta<\frac{\pi}{2}$ cases by taking the scaling limit~\eqref{scaling}, there are questions for which the non-universal subleading effects are also important. For example, the exact location of the extended boundary and the QES entail more fine-graned knowledge about the $(d+1)$-dimensional geometry near the Planck brane. In the future, it would be interesting to address them with the numerical solution~\cite{Almheiri:2019psy}. 

% Notice, however, that the numerical solutions constructed in~\cite{Almheiri:2019psy} contains only the exterior regions. To study the corrections to the HM surface, the interior solutions are also needed. 

% In our model, taking $b$ large allows us to not only scale out the effects of brane backreaction, but also make the island surfaces large and therefore applicable for membrane theory of entanglement dynamics~\cite{Jonay:2018yei,Mezei:2018jco}. While in this large $b$ limit the quantum extremal island rule still applies, and there is also an information paradox similar to that for the eternal two-sided black holes~\cite{Almheiri:2019yqk}, it would be interesting to find other double-holographic entanglement membrane models with setups in closer resemblance of the black hole information paradox. One possible starting point is to consider theories with intrinsic gravity on the brane~\cite{Chen:2020hmv,Chen:2020uac}.  

In our studies, we considered double holography models in which the brane tensions are constants~\eqref{Action}. It would be interesting to explore the possibility of deriving entanglement membrane theories from other models in double holography with more non-trivial brane profiles. For instance, one might consider theories with intrinsic gravity on the brane~\cite{Dvali:2000hr} studied in e.g.~\cite{Chen:2020hmv,Chen:2020uac}. A less ambitious, more technically-oriented direction is to use ideas from (generalised) membrane theory~\cite{Mezei:2018jco,Mezei:2019zyt,Jiang:2024tdj} to simplify the calculations in other double holography models that are otherwise hard to track analytically. 

Our BCFT model provides an approach to elucidating quantitative connections between Page curve calculations in semi-classical gravity and in chaotic many-body systems. While we specifically focused on entanglement membrane, it would be interesting to bridge other chaotic many-body system calculations of Page curves with gravity using similar constructions. One microscopic model to consider is the void formation formalism~\cite{Liu:2019svk,Liu:2020gnp}, and another toy model of interest is~\cite{deBoer:2023axh}. 

Most double holography models so far focus on the eternal two-sided black holes~\cite{Almheiri:2019yqk}. It would therefore be interesting to find explicit double holography constructions for $\emph{evaporating}$ black holes~\cite{Penington:2019npb,Almheiri:2019psf} based on the general idea laid out in~\cite{Almheiri:2019hni}. Before the Page time, the entropy of Hawking radiation is computed via a $\emph{joining\ quench}$ procedure that admits a simple membrane theory description~\cite{Mezei:2019zyt}. In Appendix~\ref{JoiningQuench}, we discuss the application of this membrane description in calculating the Page curve of Hawking radiations of an evaporating black hole before the Page time. A future challenge is to find brane solutions in double holography corresponding to evaporating black holes, and compute the full Page curves  in~\cite{Penington:2019npb,Almheiri:2019psf} from them. 

In addition to double holography, another approach of geometrising the entanglement island in $d=2$ is to perform a partial dimensional reduction of 3d gravity to JT gravity~\cite{Verheijden:2021yrb}. In this case, the entanglement entropy of Hawking radiation is computed by the geodesics in BTZ black hole backgrounds outside the horizon. In~\cite{Jiang:2024tdj}, a generalised membrane theory for 2d CFT holographically dual to BTZ is proposed, where it was shown that the geodesics outside the BTZ horizon in the scaling limit correspond to membranes with $v_B=1$ slope. It would be interesting to flesh out a generalised membrane theory for the model~\cite{Verheijden:2021yrb}. 

\section*{Acknowledgments}
We are deeply grateful to M. Mezei for countless helpful conversations and for reading the manuscript. We also thank Juan Hernandez, Ayan Patra, Julio Virrueta, Jorge E. Santos, and especially Marcos Riojas and Shan-Ming Ruan for discussions. HJ acknowledges hospitality at the University of Bristol during the course of this work, supported by UK Research and Innovation (UKRI) under the UK government’s Horizon Europe guarantee (EP/Y00468X/1). HJ is partially supported by Lady Margaret Hall, University of Oxford. AT acknowledges support from UK Engineering and Physical Sciences Research Council (EP/SO23607/1). MB acknowledges support from UK Research and Innovation (UKRI) under the UK government’s Horizon Europe guarantee (EP/Y00468X/1). There is no underlying data associated with this work.

% Double holography geometrizes the QES to ordinary HRT surfaces only in the holographic bulk matter limit. It would therefore be interesting to understand if one can incorporate quantum corrections to entanglement membranes without invoking double holography. 

\appendix

\section{Page curve of evaporating black holes in $d> 2$ before the Page time}\label{JoiningQuench}
\begin{figure}[htbp]
\centering
\includegraphics[width=.77\textwidth]{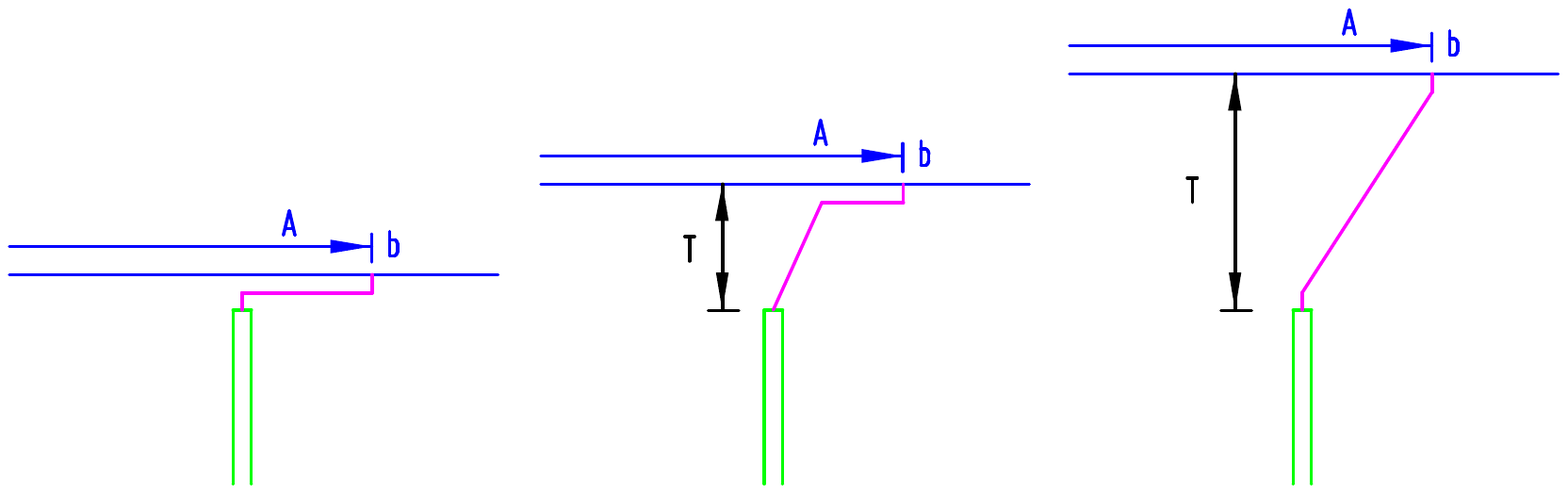}
\qquad
\caption{Entanglement membrane for a joining quench studied in~\cite{Mezei:2019zyt}. The tensionless Cardy brane is depicted as green in each picture. The reservoir we are collecting the Hawking radiation is $[b,+\infty)$. The entanglement membrane is a minimal curve starting from the boundary of the radiation region at $x=b$ on the $t=T$ slice, and ending on the Cardy brane. In the left figure we plotted for $T=0$, during which the membrane is horizontal and computes the entropy $s_\text{th} \vol(\p A) b$; in the middle figure we showed the membrane at $v_B T\leq b$, which contains horizontal segment as well as a line with slope $v_B$, and also computes the entropy $s_\text{th} \vol(\p A) b$; in the right figure, the entanglement membrane is a straight line with $v=\frac{b}{T}$ when $v_B T\geq b$, the entropy it computes is $s_\text{th} \vol(\p A)T\mathcal{E}\Big(\frac{b}{T}\Big)$. \label{fig:JoiningQuenchCartoon}}
\end{figure}

In~\cite{Almheiri:2019hni}, it was pointed out that for an evaporating AdS$_d$ black hole coupled to a flat $d$-dimensional bath, the coupling process at $T=0$ leads to a joining quench dynamics in the double holographic model at $0<T<t_P$.\footnote{The original idea in~\cite{Almheiri:2019hni} is laid out in the $d=2$ JT black hole, in which case the double holographic joining quench takes place in AdS$_3$~\cite{Ugajin:2013xxa,Shimaji:2018czt}. We expect the joining procedures of the gravitating system and the bath to be qualitatively the same in $d>2$. Like in the $d=2$ case~\cite{Almheiri:2019hni}, here in $d>2$ we also assume the coarse-grained entropy of the black hole created from the energy pulse released in the coupling process to be much larger than the entropy of the initial state. }  In this process, the HRT surface that ends on the infalling tensionless Cardy brane in the $(d+1)$-dimensional bulk computes the entropy of radiation of the evaporating black hole. Precisely the same problem was addressed in~\cite{Mezei:2019zyt} for joining quench in $d>2$ from an entanglement membrane point of view, see Figure~\ref{fig:JoiningQuenchCartoon}. The resulting entanglement entropy computed by this membrane configuration is 
\begin{equation}
\label{JoiningQuenchEE}
    S(\mathfrak{R})=s_\text{th} \vol(\p A)
    \left\{
    \begin{aligned}
        &b,&& v_B T\leq b\\
        &T\mathcal{E}\Big(\frac{b}{T}\Big)&& v_B T\geq b
    \end{aligned}
    \right.
\end{equation}
At timescale much larger than $b$, we expect $S(\mathfrak{R})\approx s_\text{th} \vol(\p A)v_E T$. 

\begin{figure}[htbp]
\centering
\includegraphics[width=.55\textwidth]{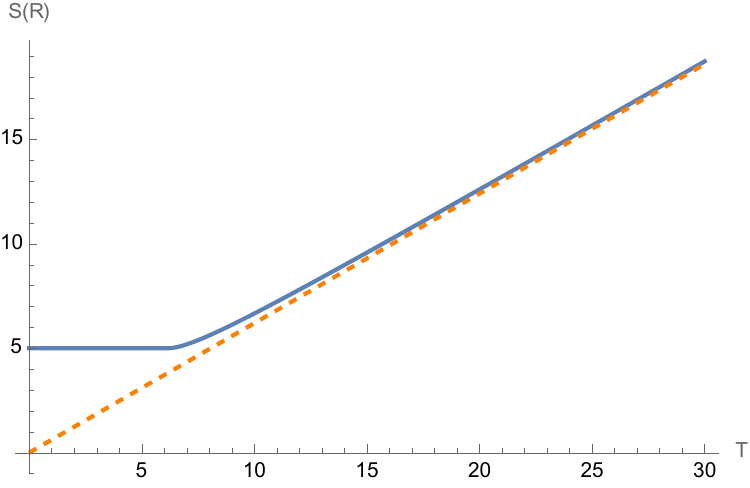}
\qquad
\caption{Entanglement entropy of Hawking radiation for an evaporating black hole before the Page time. Here we heuristically plotted with the membrane tension of AdS$_{d+1}$-Schwarzschild black brane in $d=4$, with $b=5$. The dashed orange line is with slope $v_E$. \label{fig:JoiningQuench}}
\end{figure}

Like in the eternal black hole case in the main text, in $d>2$ we would expect the HRT surface to receive corrections from the Planck brane. To heuristically illustrate \eqref{JoiningQuenchEE}, however, we can plug in the membrane tension for AdS$_{d+1}$-Schwarzschild black brane~\cite{Mezei:2018jco}. When $v_B T\geq b$ we have 
\begin{align}
    S(\mathfrak{R})=s_\text{th} \vol(\p A)\ T\frac{v_E }{\Big(1-\big(\frac{b}{T}\big)^2\Big)^\frac{d-2}{2d}}
\end{align}
See Figure~\ref{fig:JoiningQuench} for a plot of $S(\mathfrak{R})$ with respect to $T$. 

At $T>t_P$, the HRT surface ends on the Planck brane and does not depend on the tensionless Cardy brane anymore~\cite{Almheiri:2019hni}. To compute the latter half of the Page curve, one needs to find the brane profile for the evaporating black hole in $d$-dimensions.

\bibliographystyle{JHEP}
\bibliography{biblio.bib}

\end{document}